\documentclass[a4paper,fleqn]{cas-dc}

\usepackage[authoryear,longnamesfirst]{natbib}
\usepackage{algorithm}
\usepackage{algorithmicx}
\usepackage{algpseudocode} %

\usepackage[left]{lineno}

\usepackage{subcaption}

\DeclareMathOperator*{\minimize}{minimize}

\renewcommand{\cite}{\citep}
\graphicspath{{basics/figures/}}

\begin{document}
\let\WriteBookmarks\relax
\def\floatpagepagefraction{1}
\def\textpagefraction{.001}
\shorttitle{Grid-Constrained Smart Charging of Large EV Fleets}
\shortauthors{I. Kuvvetli et~al.}

\title [mode = title]{Grid-Constrained Smart Charging of Large EV Fleets:  \\ Comparative Study of Sequential DP and a Full Fleet Solver}
\tnotemark[1]

\tnotetext[1]{This research was supported by the Swedish Energy Agency. }

\author[1,2]{Ipek Kuvvetli}[type=editor,
	auid=000,bioid=1,
	orcid=0000-0003-2517-0631]
\cormark[1]
\ead{ipek.kuvvetli@vti.se}
\ead[url]{https://liu.se/en/employee/ipeku88}

\affiliation[1]{organization={Swedish National Road and Transport Research Institute (VTI)},
	city={Linköping},
	postcode={581 95},
	country={Sweden}}

\author[2]{Christofer Sundström}
\ead{christofer.sundstrom@liu.se}
\author[1,2]{Sogol Kharrazi}[%
]
\ead{sogol.kharrazi@vti.se}

\affiliation[2]{organization={Linköping University, Department of Electrical Engineering},
	addressline={Campus Valla},
	postcode={581 83},
	city={Linköping},
	country={Sweden}}

\author[2]{Erik Frisk}
\ead{erik.frisk@liu.se}

\cortext[cor1]{Corresponding author}

\begin{abstract}
	This paper presents a comparative optimization framework for smart charging of electrified vehicle fleets. Using heuristic sequential dynamic programming (SeqDP), the framework minimizes electricity costs while adhering to constraints related to the power grid, charging infrastructure, vehicle availability, and simple considerations of battery aging. Based on real-world operational data, the model incorporates discrete energy states, time-varying tariffs, and state-of-charge (SoC) targets to deliver a scalable and cost-effective solution. Classical DP approach suffers from exponential computational complexity as the problem size increases. This becomes particularly problematic when conducting monthly-scale analyses aimed at minimizing peak power demand across all vehicles. The extended time horizon, coupled with multi-state decision-making, renders exact optimization impractical at larger scales. To address this, a heuristic method is employed to enable systematic aggregation and tractable computation for the Non-Linear Programming (NLP) problem. Rather than seeking a globally optimal solution, this study focuses on a time-efficient smart charging strategy that aims to minimize energy cost while flattening the overall power profile. In this context, a sequential heuristic DP approach is proposed. Its performance is evaluated against a full-fleet solver using Gurobi, a widely used commercial solver in both academia and industry. The proposed algorithm achieves a reduction of the overall cost and peak power by more than 90\% compared to uncontrolled schedules. Its relative cost remains within 9\% of the optimal values obtained from the full-fleet solver, and its relative peak-power deviation stays below 15\% for larger fleets.
\end{abstract}


\maketitle

\section{Introduction}
Rapid growth in the adoption of electric vehicles (EV) is significantly changing the modern transportation sector. This trend is closely related to increasing environmental concerns, including reductions in greenhouse gas emissions, the depletion of fossil fuel resources, and the mitigation of urban noise pollution \cite{article, article2, SACHAN2020102238}. On a broader perspective, building a sustainable city future requires cutting emissions by adopting alternative mobility solutions, with wide fleet transitions at the center. In this context, electrified public bus fleets are crucial: They provide reliable, affordable service while delivering significant life-cycle emissions reductions. Across numerous countries, municipal governments encourage the deployment of battery-electric buses (BEBs) through measures such as financial incentives, consumer subsidies, and petroleum taxes \cite{DUAN2023104175, HE2022103437, Electricity2020, PowerCircle2025_elbussar}. In Sweden, the number of BEBs increased from about 1,140 in 2023 to 1,675 in traffic after the first quarter of 2025 \cite{Larsson2024_elbussar}. According to results on zero-emission city bus shares between 2021 and 2024 in Europe \cite{Molliere2025HalfOfNewEUBuses}, the Netherlands is a pioneer in shifting away from petroleum-powered transport, with new diesel buses accounting for less than 1\% of recently registered city buses. This transition is followed by the Nordic countries, then a few Eastern and Southern European countries.
As electrified bus fleets gain notable attention and market share worldwide, this level of fleet electrification drives high electrical power demand, which directly poses significant challenges for the electric power grid. When numerous BEBs are charged simultaneously in an uncoordinated manner, the resulting aggregated demand may place substantial stress on the grid, potentially causing power supply imbalances, increased energy losses, and voltage instability \cite{ZAFAR2023108861, BARUA2025111359, DAINA201736, su141912077}.

\begin{figure}
	\centering
	\includegraphics[width=1\columnwidth]{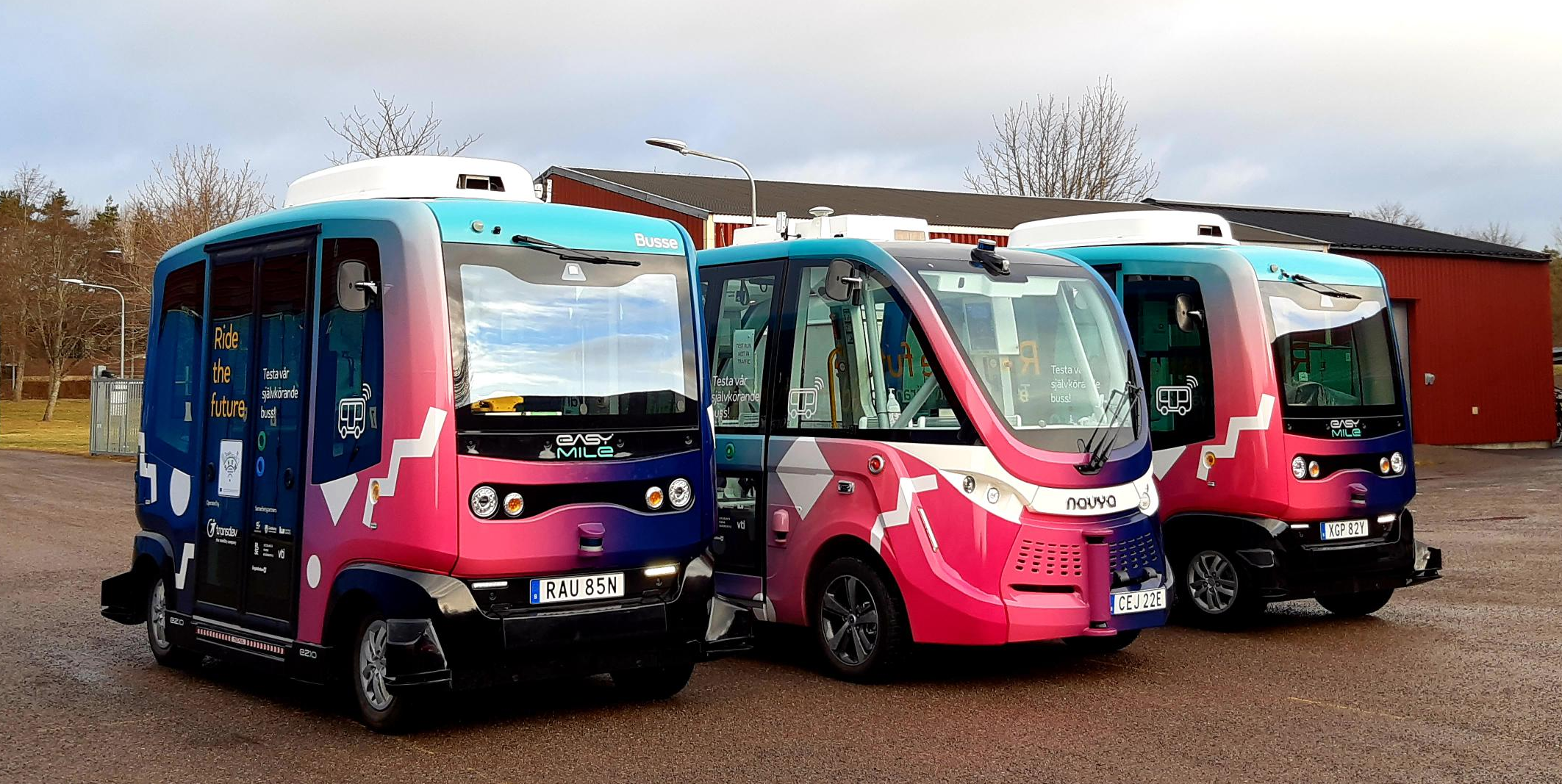} %
	\caption{Electric shuttles scheduled for intelligent charging.}
	\label{fig:fleet}
\end{figure}

Over the last decade, with the increase in BEBs,  the implementation of smart charging algorithms has become increasingly important \cite{8646425, 7287792}. These systems perform optimized charging schedules and dynamic load management \cite{6168819}, ensuring that EV charging activities are aligned with the grid capacity and user requirements \cite{JIAN2014508, 5212041, 5589729, su152216073}. By leveraging real-time data acquisition, advanced control algorithms, and communication technologies, intelligent charging offers a robust strategy for supporting large-scale EV deployment while preserving the reliability and efficiency of the power grid \cite{SOARES2014293, 5986769, KANG2013767}. Among smart-charging strategies, approaches that minimize costs \cite{VALENTINE201110717, 9259209, article5, DALLINGER20123370} and reduce grid peak demand \cite{TUCHNITZ2021116382, 9259209, MKAHL2017177} are common objectives. Additionally, computational efficiency becomes crucial for real-world implementations as the fleet sizes grow and the planning horizon expands.

Electricity prices in most European countries are set at an hourly rate \cite{nordpool2025}. The current electricity tariff imposes a monthly charging plan that penalizes the maximum power drawn over the month \cite{TekniskaVerken_Elnat2025}. Overall, this paper focuses on monthly cost optimization by applying an upper bound on grid power draw over a relatively long time horizon, in contrast to recent studies, given in Section~\ref{sec:relativeworks}. The study highlights the proposed methods' features in terms of accuracy, scalability, applicability, and computational complexity at the fleet level by benchmarking against a commercial general purpose solver.

\begin{figure*}
	\centering
	\includegraphics[width=\textwidth]{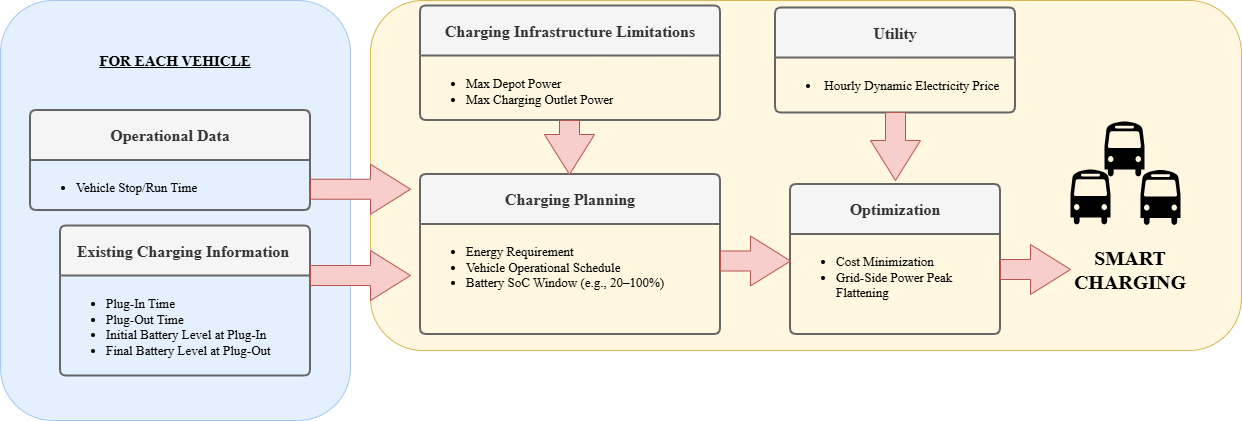}
	\caption{Proposed smart charging general configuration.}
	\label{fig:general_structure}
\end{figure*}

The objective is to design an algorithm to optimize depot charging planning of fleets of electrified vehicles, accounting for energy costs and grid constraints. The optimization problem considers peak-power tariff and energy-cost minimization, while the driving missions are to be fulfilled.
The potential of the smart charging algorithm is to be found, and thereby, all operational information is assumed to be known.
Even though the optimization is based on the knowledge of the future operation, the computational complexity increases since the power tariff depends on the peak power for one month.
Therefore, a computationally efficient framework combining optimization and heuristics in a novel way is to be developed to be able to handle large fleets of BEBs.
To evaluate the performance of the proposed methodology, it is compared against a Linear Programming (LP) structure derived from the original Non-Linear Programming (NLP) problem and implemented in a commercial optimization software.
Furthermore, the feasibility of the developed method is illustrated using a real case study with three autonomous shuttles (see Figure~\ref{fig:fleet}).
In Figure~\ref{fig:general_structure}, the main overall architecture of the proposed smart-charging technique is presented.
In this flow, the energy requirement of the shuttles in the charging planning box is directly taken from the authors' previous study \cite{Kuvvetli2025EnergyShuttles}.  To show the benefits of the methodology, the case study is up-scaled to 50 vehicles. Since the method is generic, the framework can be used for city buses, shuttles, and other fleets, including electric trucks, beyond mini-shuttles.

\subsection{Problem Formulation}

The aim of this paper is to investigate the upper bound of intelligent charging on a fleet level within an optimal planning framework that accounts for peak power tariffs. As the time horizon expands, the time complexity per vehicle increases linearly. For intelligent charging methodologies, algorithmic complexity is correlated with the number of vehicles in a fleet. In particular, the original formulation yields exponential growth in time complexity with respect to the number of vehicles in the charging plan. This study works on three major principles:

\begin{itemize}
	\item to derive an upper bound on the grid power of fleets to flatten peak power demand,
	\item to propose a practical method to handle the exponential computational complexity, which is a result of increasing fleet size,
	\item to develop a heuristic scalable optimization, whose performance is benchmarked against a commercial fleet solver.

\end{itemize}

\subsection{Outline}
The paper is structured as follows: Section~\ref{sec:relativeworks} presents related works in literature.  In Section~\ref{sec:SDP}, a heuristic method for solving the smart charging optimization problem is described. In Section~\ref{sec:gurobi}, it is followed by the description of the NLP problem and how it is converted into the LP structure and solved accordingly via Gurobi software.
In Section~\ref{sec:casestudy}, the system of the shuttles used in the case study is described. Next, the results and conclusions are presented in Sections~\ref{sec:results} and~\ref{sec:conclusion}.

\section{Related Works}
\label{sec:relativeworks}

Several recent studies investigate reinforcement learning for electric-vehicle fleet charging planning as a dynamic alternative to classical optimization \cite{en17215442,su151813553,9444352,9794744}. For instance, a proximal policy optimization-based deep reinforcement learning schedules municipal fleet operations with renewable utilization in \cite{en17215442}, and the authors in \cite{su151813553} proposed a joint charging framework that learns when and where vehicles should charge, cutting charging wait times and reducing passenger wait times. Other examples are a study focusing on state clustering with a compressed actor–critic of linear complexity, lowering total charging costs while preserving fast computation \cite{9444352}; and a decomposition-based forecasting pipeline, which achieves a high coefficient of determination, supporting demand-aware scheduling \cite{9794744}. In general, these approaches typically require substantial training data, careful design of rewards and hyperparameters, and significant computation, and they remain challenging to interpret and transfer reliably across fleets and operating conditions.

Dynamic Programming (DP) is a widely adopted approach for tackling optimal control problems. It is commonly applied to model energy usage in individual vehicles. For instance, daily charging profiles are optimized via peak-shaving and valley-filling \cite{6874180}; and vehicle-level DP is proposed, and it enables V2G operations, yielding lower electricity costs than uncontrolled charging, and it is compared with Linear Programming (LP) methods, which are not able to incorporate changes in charging efficiency \cite{en18051109}. In \cite{wevj15030090}, the authors apply the DP to optimize charging and battery preheating for an individual EV to meet the target SoC by cutting energy cost. Approximate DP (ADP) methods, such as  Least-Squares Monte Carlo-based approaches, facilitate intelligent charging under price and usage uncertainty for small to mid-size fleets over short horizons, with shown cost reductions and peak-load mitigation \cite{LEE2022119793,MAHYARI2025}. ADP cannot guarantee global optimality because it uses sample-trained value-function approximations and substitutes exact recursion with a one-step lookahead, yielding only approximate Bellman consistency. It basically requires the immediate reward plus the discounted expected value of the next state.

In literature, pure dynamic programming is usually limited to single electric vehicles or small fleets because of state-space explosion, to handle multi-state uncertainty at scale. In light of this, the authors in \cite{SKUGOR2015456} addressed the computational complexity issue at the fleet level by aggregating vehicles into a single storage entity. Given these computational complexity issues, researchers turn to heuristic approaches or commercial optimizers, such as Gurobi. In the existing literature, some models are solved using Gurobi across various case studies \cite{chawal2023optimizing, 9903610, su152216073,akaber2021milp, zhao2025wmc, alkanj2018approximatedynamicprogrammingplanning} because Gurobi-based LP and NLP optimizations offer clarity and optimality in energy management.

Unlike studies focused on single vehicles or 24-hour time windows, this study optimizes over a monthly horizon for a growing shuttle fleet. While the current data used in this paper covers mini-shuttles, the framework is generic and designed to scale to heterogeneous fleets, including, e.g., electric shuttles, city buses, or trucks. Unlike ADP, this study assumes full observability of all electric vehicles' available charging slots and energy requirements, so it does not rely on statistical information.  Therefore, this study focuses on a practical, deterministic DP approach for long-horizon planning and benchmarks it against the LP formulation, obtained from the original NLP problem, solved using Gurobi, which provides globally optimal solutions. Although the proposed method introduces discretization error, it yields solutions close to the global optimum. Unlike existing studies \cite{SKUGOR2015456}, it does not require aggregating vehicles based on energy requirements. Subject to grid power constraints, charging decisions are made sequentially. At each time step, the available power is reduced if the previous vehicle is already being charged in the same time slot. Hence, the drawable power is upper-bounded by the charging schedules of the previously optimized vehicles. In the next section, the proposed method is explained in detail.

\section{Sequential Heuristic DP Optimization Methodology}
\label{sec:SDP}

Considering the relevant literature, Dynamic Programming is commonly used in charging planning and guarantees a globally optimal solution, up to discretization. However, as mentioned before, a major drawback is that its computational complexity grows exponentially with the increased number of state variables in the dynamic system. The extended planning horizon also leads to a linear increment in elapsed time \cite{guzzella2013vehicle}. With an increasing number of vehicles in the planning problem, the number of states increases, and the computational complexity makes direct DP infeasible. To avoid this limitation, a heuristic approach combined with DP is developed to manage the complexity.

In the proposed Sequential DP configuration, the state variables are the vehicle's battery state of charge, implying that the number of states grows linearly with the number of vehicles in the fleet. However, the vehicles' charge plans are determined via a sequential decision process, which reduces the number of states to 1 at each step in the optimization sequence.  Each vehicle's grid-drawable power capacity is upper-bounded by the previous charging allocation. This simplification allows the application of DP while maintaining a practical connection to the minimum-cost principle for a long-term horizon, as required by the new power tariffs.  As a result, the proposed method offers time-efficient optimization for a large fleet, sacrificing global optimality for near-optimal results with significantly lower computational complexity. First, the cost function and related constraints are given. The proposed DP methodology is described in detail afterward.

\subsection{Cost Function and Constraints}

In this section, all formulations of the optimization problem are introduced in detail in~\eqref{eq:all_cons}. The cost function of the original problem, given in~\eqref{eq:costfunc}, is defined with an energy term based on spot market prices and a penalty for peak power usage.  In this formulation,  $c_{\text{spot},t}$ denotes the spot electricity price [EUR/kWh] at time $t$, and $c_m$ represents the peak power cost coefficient [EUR/kW]. It is noted that $c_{\text{spot},t}$ denotes the price per energy unit, whereas $c_m$ denotes the price per power unit. In this formulation, $P_{k,t}$ is the charging power [kW] of shuttle $k$ at time $t$. The first part of the cost function accounts for total energy consumption up to the end time $T$ and for all shuttles $K$, with $\Delta \tau$ denoting the duration of each time step. The second part aggregates all time-dependent power values across all vehicles and sums them across each time period to identify the maximum power within the time horizon. This part penalizes the maximum value over the entire monthly horizon to reduce peak power demand. The time step duration, denoted by $\Delta \tau$, may vary depending on the computation interval, as operational activities do not always align perfectly with discrete hourly boundaries (e.g., 01:00--02:00--03:00). As a result, $\Delta \tau$ may take non-integer values (e.g., 0.58 hours). The complete optimization problem can be stated as
\begin{subequations}
	\label{eq:all_cons}
	\begin{align}\hspace*{-10mm}
		\minimize_{P_{k,t},\ \text{SoC}_{k,t},\ P_{\text{max, tariff}}}\, \quad
		                       & \sum_{t=1}^{T} \sum_{k=1}^{K} P_{k,t} \, c_{\text{spot},t} \, \Delta \tau_t
		+ c_m \, \max_{t} \sum_{k=1}^{K} P_{k, t}  \label{eq:costfunc}
		                       & \qquad                                                                                             \\
		\text{subject to}\quad & \text{SoC}_{k,t+1} = \text{SoC}_{k,t} + \frac{P_{k,t} \, \Delta \tau_t}{C_k}  \nonumber            \\
		                       & \quad\quad\quad\quad-\Delta \text{SoC\_op}_{k,t}, \label{eq:soc_transit}                           \\
		                       & \text{SoC}_{\min} \le \text{SoC}_{k,t} \le \text{SoC}_{\max}, \label{eq:soc_bounds}                \\
		                       & \text{SoC}_{k,0} = 100\%,\label{eq:soc_init}                                                       \\
		                       & \left| \text{SoC}_{k,T} - \text{SoC}^{\text{target}}_k \right| \le \epsilon, \label{eq:soc_target} \\
		                       & P_{k,t} = 0 \ \text{if}\  \Delta \text{SoC\_op}_{k,t} > 0 \ \text{or}\label{eq:availability}       \\
		                       & \quad\quad\quad\quad\quad \sigma_{k,t} < 0.5~\text{h},\nonumber                                    \\
		                       & 0 \le P_{k,t} \le P_{\text{max}}, \label{eq:power_bounds}                                          \\
		                       & \sum_{k=1}^{K} P_{k,t} \le P_{\text{grid,max}} \label{eq:grid_bounds}
	\end{align}
\end{subequations}
where constraints are defined $\forall\,k\in\{1,\dots,K\},\ \forall\,t\in\{1,\dots,T\}$. The constraints are given in \eqref{eq:soc_transit}, \eqref{eq:soc_bounds}, \eqref{eq:soc_init}, \eqref{eq:soc_target}, \eqref{eq:availability}, \eqref{eq:power_bounds}, and \eqref{eq:grid_bounds}. They are included to ensure that the charging profiles for each shuttle $k$ at time $t$ are valid and feasible.
The state of charge (SoC) of shuttle $k$ at time $t$ depends on its charging or operational status, dynamically modeled by Equation~\eqref{eq:soc_transit}, and $C_k$ is the battery capacity [kWh]. To control battery health degradation, SoC needs to be contained in an interval [$\text{SoC}_\text{min}$, $\text{SoC}_\text{max}$]. This is guaranteed by including a constraint ~\eqref{eq:soc_bounds}, where $\text{SoC}_\text{min}$ and $\text{SoC}_\text{max}$ denote lower and upper bounds of the battery SoC of each vehicle, ensuring an appropriate SoC range to keep the battery health. The battery initial SoC condition is fixed to 100\% in~\eqref{eq:soc_init}. 	In~\eqref{eq:soc_target}, the target SoC, ${SoC}^{\text{target}}_k$, is the battery SoC level that is aimed for at the end of the month. The tolerance $\epsilon$ avoids infeasibility caused by discretization.

In \eqref{eq:availability}, the term $\Delta \text{SoC\_op}_{k,t}$ denotes the SoC decrement due to vehicle operation. $\Delta \text{SoC\_op}_{k,t}$ is related to the energy consumption at time index $t$ since the vehicle is in use. In this constraint, $\sigma_{k,t}$ shows the time duration when the vehicle is not in operation, and there is a charging opportunity. Considering the real-world operations, $\sigma_{k, t}$ is constrained to max 0.5 hours. If the availability is less than half an hour, charging is not allowed, and $P_{k,t} = 0$. Equation~\eqref{eq:availability} directly affects~\eqref{eq:soc_transit}.

In~\eqref{eq:power_bounds}, $P_{\text{max}}$ denotes the charging station's maximum power capacity [kW] for every shuttle, considering the charging infrastructure limitation. To prevent excessive load on the power grid, the aggregated charging power drawn by all shuttles at any time step $t$ must not exceed a predefined grid capacity. In ~\eqref{eq:grid_bounds}, $P_{\text{grid, max}}$ denotes the maximum allowable total charging power drawn from the grid. Minimization of $P_{\text{max,tariff}}$ in the loss function is directly related to the second part of~\eqref{eq:costfunc}, and it corresponds to the total aggregated charging power at any time instant t. This notation is elaborated in Section~\ref{subsec:seq_Heuristic}.

The original problem~\eqref{eq:all_cons} with a large number of vehicles, $k$, the total number of states is high, and classical DP becomes impractical for larger fleets because the time complexity scales non-linearly with the number of states. Which motivates the heuristic, sequential, DP approach described next. Further, as written, \eqref{eq:all_cons} is a non-linear program due to the max-term in the cost function. However, this will be rewritten, using standard techniques, into an equivalent linear program used in the full-fleet solver.

\subsection{Sequential Heuristic DP Algorithm Design}
\label{subsec:seq_Heuristic}
Each vehicle's charging power in the optimization is upper-bounded by $P_{\text{grid, max}}$, which is a hard constraint. The optimal power drawn from the grid should be equal to, or less than, $P_{\text{grid, max}}$. Therefore, the algorithm requires discretized power levels to find the optimal peak power, due to the second term in the cost function~\eqref{eq:costfunc}. This discretization is performed using integer steps, with $P_{\text{max, tariff}}$ ranging from 0 to $P_{\text{grid, max}}$. The total power drawn from the grid cannot exceed $P_{\text{max, tariff}}$.

The terminal cost does not include peak demand charges directly; they are accounted for after optimization by evaluating the aggregate peak power across all vehicle schedules and applying the corresponding penalty externally. Within this concept, the algorithm solves the first part of~\eqref{eq:costfunc} for each discretized subset, $P_{\text{max, tariff}}$. Thereby, the linear part of the primary cost function is given by
\begin{align}
	\minimize_{P_{k,t},\ \text{SoC}_{k,t}}\, \quad
	 & \sum_{t=1}^{T} \sum_{k=1}^{K} P_{k,t} \, c_{\text{spot},t} \, \Delta \tau_t.  \label{eq:costfunc_mod}
\end{align}

The main objective of the algorithm is to decide an upper bound for peak power and minimize the energy cost. Thereby, the aggregated power at the same time instant must be equal to or lower than the selected discretized subset of $P_{\text{max,tariff}}$ given by
\begin{align}
	\sum_{k=1}^{K} P_{k,t} \le P_{\text{max,tariff}}. \label{eq:grid_bounds_tariff}
\end{align}
The corresponding peak power penalty, associated with the second part of~\eqref{eq:costfunc}, is added to the total cost for each discretized subset, and the minimum cost is determined. The proposed algorithm solves each vehicle's optimization problem sequentially to avoid exponential growth in the number of states and reduce computational complexity. Using the discrete power levels of $P_{\text{max, tariff}}$, for each subset, the decision is made sequentially for all vehicles. Due to the power already allocated to preceding vehicles, the available power limit for subsequent vehicles is reduced at the same time step. Here, the Equation~\eqref{eq:grid_bounds_tariff} constraints are modified and rewritten as
\begin{align}
	P_{k,t}  \le \min\!\bigl( P_{\text{max}},\; P_{\text{max, tariff}} - \sum_{n=1}^{k-1} P_{n,t}\bigr) = P_{k, t}^{\text{avail}}. \label{eq:rem_bounds}
\end{align}
This formulation computes the maximum available power $P_{k,t}^{\text{avail}}$ for the $k$th vehicle in the sequential approach SeqDP, where $n$ denotes the index of the previous vehicles. Thereby, $P_{k,t}^{\text{avail}}$ corresponds to the residual grid capacity for each vehicle.

In the DP framework, the \textit{cost-to-go matrix} represents the minimum cumulative cost from a given system state (a specific SoC level at a given time) to the end of the planning horizon. Let $J_t(\text{SoC}_{t})$ denote the cost-to-go matrix at time $t$ for a given SoC level. The recursive structure of the DP can be expressed as
\begin{equation}
	J_t(\mathrm{SoC}_t)
	= \min_{ P_t \le P_{\max,\mathrm{tariff}}}
	\left\{ P_t \, c_{\mathrm{spot},t} \, \Delta \tau_t + J_{t+1}(\mathrm{SoC}_{t+1}) \right\}
	\label{eq:main_prin}
\end{equation}
where  $P_t$ denotes the charging power decision at time $t$ and the term $J_{t+1}(\text{SoC}_{t+1})$ represents the future cost-to-go from the updated state. For each vehicle $k$ and time step $t$, the charging power $P_{k,t}$ is discretized into an action set $\mathcal{A}_{k,t}$ of $N$ evenly spaced power levels between $0$ and $P_{\max}$, where $N = P_{\max}/\Delta P + 1$ is determined by the chosen resolution $\Delta P$. The state-of-charge at time $t$ is denoted by $\text{SoC}_t$, and the resulting state-of-charge at time $t+1$ after applying the action $P_t$ is $\text{SoC}_{t+1}$. Since SoC transitions are not purely discrete, the resulting $\text{SoC}_{t+1}$ typically does not align exactly with the discretized SoC grid used in the DP algorithm. Under these alignment-error conditions, the cost-to-go value is interpolated from adjacent grid points. The cost-to-go matrix is computed in a backward pass starting from the end of the time horizon. At each step, the minimum cost is calculated by reaching the goal from every possible battery level. This process constructs a cost-to-go table that provides the best cost the algorithm can achieve from any given state and time. The control sequence is then recovered in a forward pass \cite{guzzella2013vehicle}.

\begin{algorithm}
	\centering
	\footnotesize
	\caption{Sequential Heuristic DP}
	\label{alg:seq-pm}
	\begin{algorithmic}[1]
		\Require Months $\mathcal{M}$; $P_{\text{max,tariff}}$ : discretized subsets of hard-constraint peak power of $\mathcal{P_{\text{grid, max}}}$; vehicles $k=\{1,2,3\}$; time grid $\mathcal{T}$ (step $\Delta \tau$);
		spot prices $\{c_{\mathrm{spot},t}\}_{t\in\mathcal{T}}$; demand-charge rate $c_m$; charger capacity $\{P_{\max}\}$.
		\Ensure For each month $m\in\mathcal{M}$: optimal peak $P_m^\star$, schedules $\{P_{k,t}^\star\}$, and minimal total cost.

		\For{$m \in \mathcal{M}$}
		\State $\text{best\_cost} \gets +\infty$,\quad $P_m^\star \gets \text{none}$
		\ForAll{$P_{\text{max,tariff}} \in \mathcal{P_{\text{grid, max}}}$}
		\State $G_t \gets P_{\text{max,tariff}}$ for all $t\in\mathcal{T}$ \Comment{residual grid capacity}
		\State feasible $\gets$ true;\quad energy\_cost $\gets 0$;\quad initialize $P_{k,t}\gets 0$
		\For{$k \in \{1,2,3\}$}
		\State Solve DP for vehicle $k$:
		\Statex \hspace{1.5em} $\displaystyle \min_{\,\{P_{k,t}\}} \sum_{t\in\mathcal{T}} c_{\mathrm{spot},t}\,P_{k,t}\,\Delta \tau$
		\Statex \hspace{1.5em} subject to $0 \le P_{k,t} \le \min\{P_{\max},\,G_t\}$\ \ $\forall t\in\mathcal{T}$
		\If{DP infeasible}
		\State feasible $\gets$ false
		\State \textbf{break} \Comment{exit vehicle loop}
		\EndIf
		\State $G_t \gets G_t - P_{k,t}$ for all $t\in\mathcal{T}$
		\State energy\_cost $\gets$ energy\_cost $+ \sum_{t\in\mathcal{T}} c_{\mathrm{spot},t}\,P_{k,t}\,\Delta \tau$
		\EndFor
		\If{not feasible}
		\State \textbf{continue} \Comment{try next $P\in\mathcal{P_{\text{grid, max}}}$}
		\EndIf
		\State demand\_cost $\gets c_m \cdot P_{\text{max,tariff}}$
		\State total\_cost $\gets$ energy\_cost $+$ demand\_cost
		\If{total\_cost $<$ best\_cost}
		\State $\text{best\_cost} \gets \text{total\_cost}$
		\State $P_m^\star \gets P_{\text{max,tariff}}$
		\State $P_{k,t}^\star \gets P_{k,t}$
		\EndIf
		\EndFor
		\State \textbf{Output} $P_m^\star$, $\{P_{k,t}^\star\}$, and $\text{best\_cost}$ for month $m$
		\EndFor
	\end{algorithmic}
\end{algorithm}
The proposed Algorithm~\ref{alg:seq-pm} illustrates the sequential heuristic DP approach in detail. The algorithm uses only 3 vehicles to provide a small-scale representation and requires a monthly time horizon.
The proposed design employs sequential decision-making with discrete power levels. The algorithm initially solves the first part of~\eqref{eq:costfunc} for each discretized subset, with a detailed representation in~\eqref{eq:main_prin}. In the inner loop, the algorithm tries to minimize energy cost for each vehicle sequentially. If there is allocated power at the same time slot in the preceding vehicle charging plan, residual grid capacity is reduced, and charging power at that slot must be equal to or less than the sequentially determined grid power limit. The total cost is not directly included as a peak-demand charge; the optimization aggregates peak power across all vehicle schedules and applies the corresponding penalty externally, thereby selecting a subset of peak power, $P_{\text{max, tariff}}$.

In this sequential approach, the vehicles are ordered. Choice of ordering affects the solution and here the vehicles are ordered from the highest to the lowest energy requirement. The method accommodates scalable fleet sizes and infrastructure constraints, while enabling real-time integration of spot-market electricity prices and grid capacity constraints. Its structure allows deployment in both centralized and distributed fleet charging scenarios.

\section{Full Fleet Solver}
\label{sec:gurobi}
The full-fleet optimization problem is a nonlinear program and the solver Gurobi \cite{gurobi} is used for benchmarking and compare with the solutions from the sequential DP approach. Hence, the joint solution obtained by this approach is globally optimal, planning all vehicles simultaneously rather than sequentially, as in SeqDP. The cost function described in Section~\ref{sec:SDP} is the same, but the applied model is constructed by considering all states simultaneously when using the fleet solver to find the global optimal solution.

The fleet solver directly employs the equation \eqref{eq:grid_bounds} using a single decision variable, thereby avoiding sequential decisions. Thus, $P_{\text{grid,max}}$ is always at least as large as the total charging power at time $t$.  In the NLP formulation of the original problem, the nonlinear part is directly derived from the max operator in \eqref{eq:costfunc}. To avoid this, a standard reformulation approach is used, transforming the problem into an LP problem by upper-bounding the total aggregating power over the entire time series. For this, replace the max term in the cost function with the variable $P_{\text{max,tariff}}$, and instead introduce the linear constraint
\begin{align}
	\sum_{k} P_{k,t} & \le P_{\text{max,tariff}}, \qquad \forall t,
	\label{eq:LP_reform}
\end{align}
in the problem and optimize also over $P_{\text{max,tariff}}$ as a decision variable. The Charging availability, assumed known or accurately estimated, is encoded by the parameter $u_{k,t}$, and the solution is constrained by this parameter. Thus, the decision variables are $x=[P; P_{\text{max,tariff}}]$, where $P=\{P_{k,t}\}_{k=1..K,\,t=1..T_k}$ and $ P_{\text{max,tariff}}$ is the maximum peak-aggregated power for the entire month.
The applied general LP formulation is
\begin{align}
	\minimize_{x}\quad & c^\top x                                          \\
	\text{s.t.}\quad   & A x \le b,\ \ell \le x \le u,\ x_i \in \mathbb{R}
	\label{eq:power_gate_grid_cap}
\end{align}
and the operational constraint related to charging availability is represented by
\begin{align}
	u_{k,t} & =
	\begin{cases}
		1, & \text{if } \Delta\tau_{k,t}\ge 0.5~\mathrm{h}\ \text{and}\  \Delta \mathrm{SoC\_op}_{k,t}<0, \\
		0, & \text{otherwise},
	\end{cases}.
	\label{eq:availability2}
\end{align}
Charging power is constrained by vehicle availability, and the grid and peak constraints as
\begin{align}
	0 \le P_{k,t}               & \le u_{k,t}\,P_{\max}, \qquad \forall k,t, \nonumber \\
	0 \le P_{\text{max,tariff}} & \le P_{\text{grid,max}}.
	\label{eq:power_gate_grid_cap2}
\end{align}
The optimal solution is obtained using linear programming by solving in continuous time. The notation of $P_{\text{max,tariff}}$ corresponds to the upper-bound aggregated grid power, which is aimed for, at the end of the solution.

LP problem-solving is generally fast, however due to the cumulative SoC characteristics a nonlinear time complexity is inherently included to the LP. The LP solver satisfies \eqref{eq:soc_transit}, \eqref{eq:soc_bounds}, \eqref{eq:soc_init}, and \eqref{eq:soc_target}.
Consider the constraint \eqref{eq:soc_transit} where SoC is a time-dependent variable that evolves dynamically as

\begin{equation}
	\text{SoC}_{k,t} = \text{SoC}_{k,0} +\sum_{m=0}^{t-1} \frac{P_{k,m} \, \Delta \tau_m}{C_k} - \sum_{m=0}^{t-1}\Delta \text{SoC\_op}_{k,m}.
	\label{eq:cumulative_soc_traj}
\end{equation}
Equation \eqref{eq:cumulative_soc_traj} gives the cumulative SoC structure. Instead of taking $\text{SoC}_{k,t}$ as an explicit optimization variable, the SoC computes the sum of all previous charging contributions and operational energy consumption. For a better understanding, let's consider the upper constraint of SoC in equation \eqref{eq:soc_bounds}, the formulation becomes
\begin{equation}
	\sum_{m=0}^{t-1}
	\frac{P_{k,m} \, \Delta \tau_m}{C_k}
	\le
	\text{SoC}_{max}
	-
	\text{SoC}_{k,0}
	+
	\sum_{m=0}^{t-1}\Delta \text{SoC\_op}_{k,m}.
	\label{eq:cumulative_upper_bound}
\end{equation}
The cumulative formulation \eqref{eq:cumulative_soc_traj} leads to a constraint at time $t$ that depends on the entire charging history, which yields quadratic computational complexity in computing the SoC used in the constraints.

\section{Case Study}
\label{sec:casestudy}

In this section, the feasibility of the proposed SeqDP framework is evaluated through a case study involving autonomous shuttles. Each shuttle is modeled and planned individually, with a planning horizon of 1 month.  This study focuses on two case studies: one is a small-scale fleet comprising three shuttles currently operating around campus at Linköping University; the other scales the algorithm to 50 vehicles. The current fleet includes 3 shuttles; it is expected to increase, and this study needs to be scalable to be implemented in the large-scale municipality's projects. To ensure long-term viability, the proposed methodology is both comprehensive and scalable, enabling application from individual vehicles to large-scale shuttle fleets. Smart charging directly optimizes charging schedules based on battery level and operational requirements, such as route planning.  Shuttle operations, which require high energy without interruptions, require the upper bound of SoC to be set to 100\%, and 20\% is considered the lower bound for battery health \cite{Oloyede2024_SOCReview}. It is also noted that each charging spot's power, $P_{\text{max}}$, is limited to 11 kW. The ${SoC}^{\text{target}}_k$ is selected to 100. The peak power penalty during winter $c_m$ is 43 SEK/kW, corresponding to around 4 EUR/kW  \cite{TekniskaVerken_Elnat2025}. The term $c_{\text{spot},t}$ is the spot price [EUR/kWh] historically taken from the energy operator's database \cite{nordpool2025}.

\subsection{Small-scale Fleet Description and Constraints}

\begin{figure}
	\centering
	\includegraphics[width=8 cm]{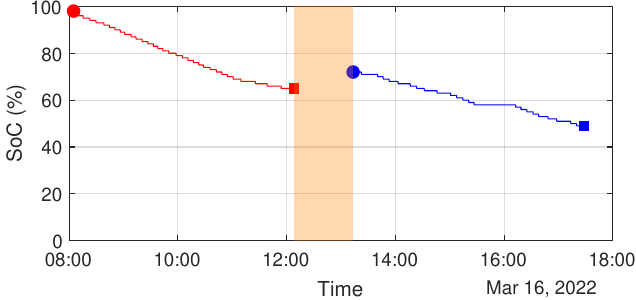}
	\caption{Detection of uncontrolled charging sessions from the difference of the previous operation's terminal SoC (red) and the following operation's initial SoC value (blue).}
	\label{fig:charge-real}
\end{figure}

\begin{figure}
	\centering
	\includegraphics[width=8 cm]{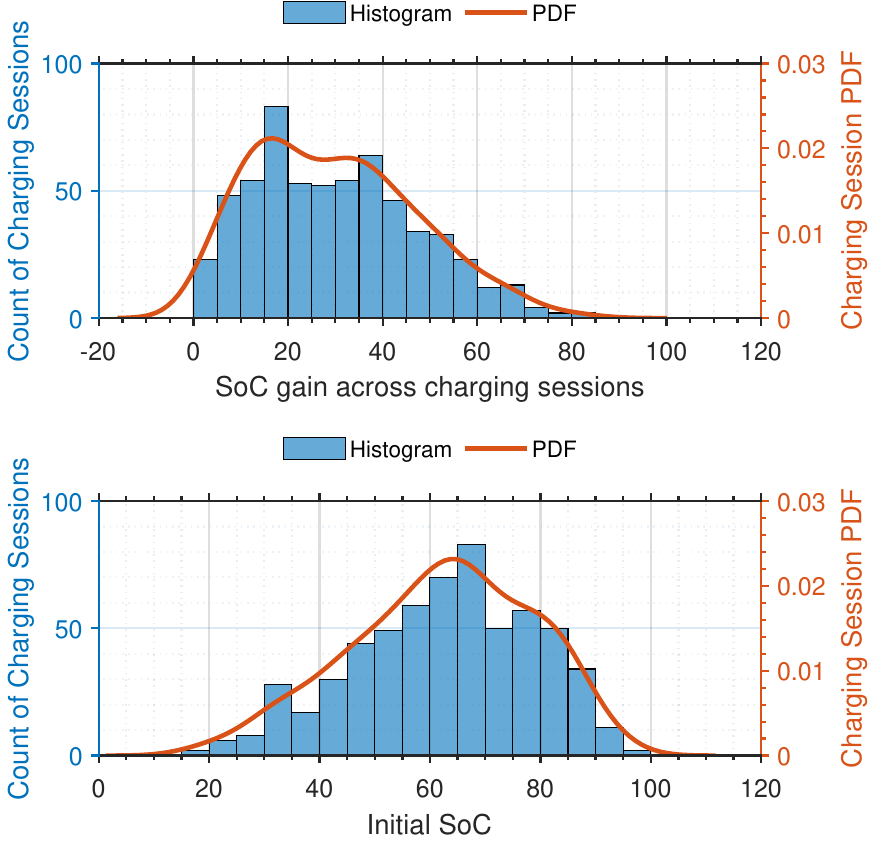}
	\caption{Distribution of SoC gain and initial SoC across charging sessions between 2020 and 2023.}
	\label{fig:soc_inc}
\end{figure}

The first case study focuses on a small-scale fleet of three shuttles performing nearly identical operations. In our case, historical data from uncontrolled charging sessions is unavailable. However, data from the vehicle operation is accessible, and therefore charging schedules are derived by processing operational data files to detect SoC gain over time. These files only provide operational timestamps and corresponding battery levels, without details on actual charging times, power levels, or charging station efficiency. By assuming a constant charging power of 5 kW for the shuttles, it is possible to distinguish between actual charging time and idle time, e.g., waiting time at the garage. The proposed method is generic; however, in this case study, fast charging is neglected, and $P_{\text{max}}$ is set to 11 kW. Each shuttle is equipped with a dedicated charging outlet. The extraction process for charging sessions is illustrated in Figure~\ref{fig:charge-real}, where the orange-filled box shows the initial and terminal plugged-in times of the shuttle, along with the corresponding SoC levels and the resulting energy requirement.  Uncontrolled charging primarily occurs between 17:00 and 07:00. Furthermore, Figure~\ref{fig:soc_inc} illustrates that the SoC gain is rarely 80-90\%.

The instantaneous power drawn from the grid is limited to $P_{\text{grid,max}}$ of 15 kW. The grid power limit depends on the number of vehicles and the aim of reducing instantaneous power demand. The maximum grid power $P_{\text{grid, max}}$ is discretized up to 15~kW in unit using integer steps, resulting in values ranging from 0 to 15~kW with a resolution of 1~kW. Based on this discretization, 16 different $P_{\text{max, tariff}}$ values are generated.  In the current three-shuttle operation, all vehicles' operations are very similar, but data acquisition is partially disabled in 2 shuttles. Thereby, the two vehicles’ schedules are offset by one hour from the base shuttle, and the required state of charge decreases slightly with each successive shuttle, about a 0.1 step down per shuttle, e.g., 0.9 then 0.8. It is also considered that the actual SoC resolution is 1\%.

\subsection{Large-scale Fleet Description and Constraints}

In the second scenario, this large-scale optimization is intended to cover a fleet of EVs operating at different times of day, with varying energy requirements. Consequently, large-scale fleet design necessitates a non-uniform distribution of energy consumption over differentiated time intervals. Unlike the relatively uniform three-shuttle schedules, more realistic fleet operations with up to 50 shuttles are synthesized using a randomized procedure. Beginning from the baseline of vehicle 1, up to 49 additional operational schedules are created by shifting all start and end times forward by a random integer number of hours uniformly selected from 0 to 10. A 10-hour random time shift is chosen with reference to current mini-shuttle operations so that, in almost every hour, at least some vehicles are operating. To model the energy consumption variation of the vehicles, for each trip, the SoC decrement is independently scaled by a uniformly sampled scale factor in the range 0.5–1.2, and the resulting SoC drops are limited to remain within the standard battery limit. Scale factors larger than 1.2 trigger immediate charging during short breaks, which is not an observed behavior in the collected data;  therefore, an upper limit of 1.2 is used. A standard battery capacity is assumed for all vehicles, so differences in energy consumption across vehicles are induced solely by the randomized SoC scaling, rather than by capacity heterogeneity.

Unlike the earlier setup with a fixed grid limit of $P_{\text{grid,max}} = 15 \  \text{kW}$, the 50-vehicle case uses $P_{\text{grid,max}} = 80 \ \text{kW}$, selected empirically by ensuring that the algorithm is feasible for that criteria. Because the computational time increases with the admissible discretized power subset, the previous restriction $P_{\text{max,tariff}} \le 15 \ \text{kW}$ is replaced by
$65 \le P_{\text{max,tariff}} \le 80 \ \text{kW}$, which narrows the search space while still allowing a near-optimal solution to be identified. For the cases with 10 and 30 vehicles, the considered power ranges are 5--20 kW and 35--50 kW, respectively. Each set contains 16 candidate values. This setup enables comparison on a common baseline.

\section{Results}
\label{sec:results}
This section illustrates the two case-study results using the SeqDP and LP solver and delves into both limitations and advantages of the proposed SeqDP. First, this section discusses how SeqDP and LP solvers efficiently reduce peak power and overall cost over a month compared to uncontrolled charging profiles. After that, the comparison of both smart algorithms is completed in terms of the relative error and computational complexity. The methodology is applied to different-sized fleet scenarios to investigate robustness and feasibility.

\subsection{Small-scale Fleet Experiment}

\begin{figure}
	\centering
	\begin{subfigure}[b]{0.48\textwidth}
		\centering
		\includegraphics[width=\linewidth]{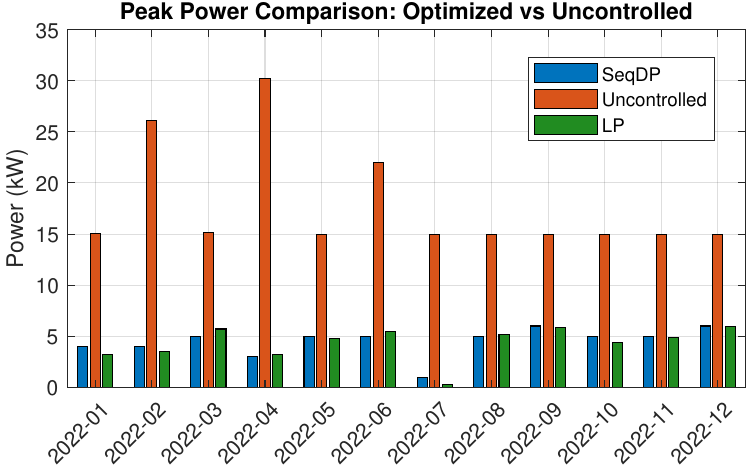}
		\caption{Smart vs uncontrolled charging grid power for 3 shuttles.}
		\label{fig:Pgrid_cmp}
	\end{subfigure}
	\hfill
	\begin{subfigure}[b]{0.48\textwidth}
		\centering
		\includegraphics[width=\linewidth]{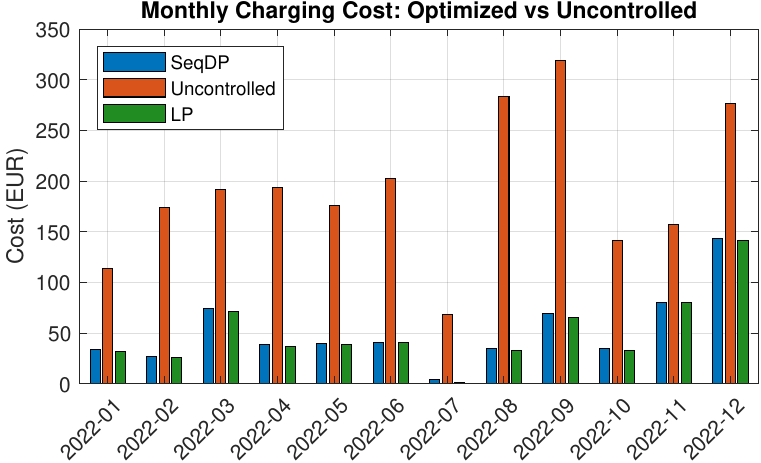}
		\caption{Smart vs uncontrolled charging cost for 3 shuttles.}
		\label{fig:Smart_uncontrolled_cost}
	\end{subfigure}

	\caption{Smart vs uncontrolled charging comparison for 3 shuttles.}
	\label{fig:smart_vs_uncontrolled_3shuttles}
\end{figure}

\begin{figure}
	\centering
	\includegraphics[width=0.45\textwidth]{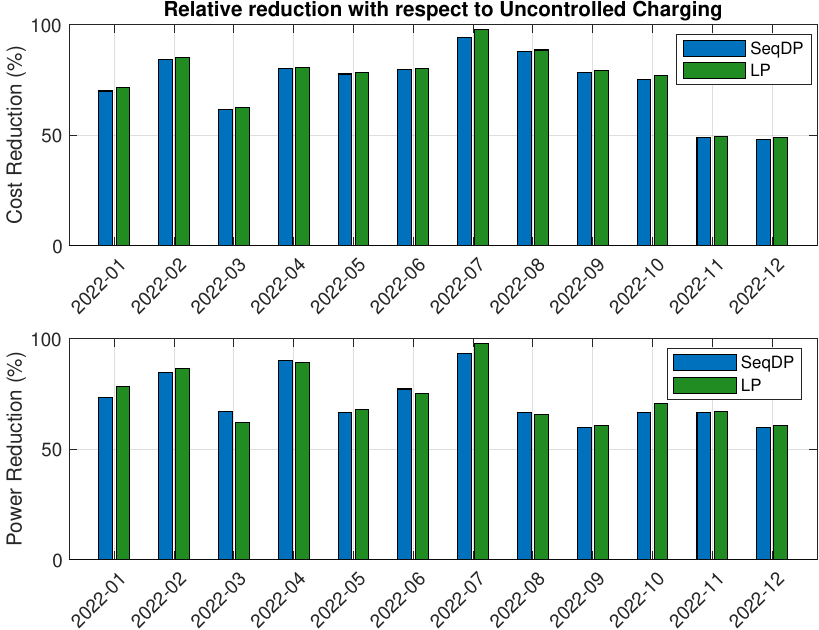}
	\caption{Relative cost and power reduction with respect to the uncontrolled charging case for 3 shuttles.}
	\label{fig:Relative_reduction_new}
\end{figure}

The simulation results compare smart charging strategies, SeqDP, and LP formulation, with uncontrolled charging profiles, highlighting significant improvements in both cost reduction and grid peak shaving (see Figure~\ref{fig:Pgrid_cmp} and Figure~\ref{fig:Smart_uncontrolled_cost}) for the 3-shuttles case. In Figure~\ref{fig:Pgrid_cmp}, it is seen that the actual total charging grid power is increased to 30 kW, while this study assumes it is 15 kW. This difference is rare and is caused by charging for a limited time between operations during a day. Smart charging at the fleet level introduces time-complexity issues as a drawback compared to uncontrolled charging. Regarding battery degradation, complete depletion is avoided, with the maximum and minimum states of charge maintained at 100\% and 20\%, respectively. As shown, the costs optimized using SeqDP are closely aligned with the globally optimal results obtained by the LP solver. Similarly, the monthly peak power values selected by SeqDP are comparable to those from the LP solver, though the discretization effects in SeqDP are evident. For the small-scale experiments, the LP solver is preferable. However, the large-scale implementations described in the next section make the LP solver impractical due to its computational complexity, even though it provides optimal solutions rather than close results as in SeqDP. As an example, July 2022 features relatively few operations, and the gap in grid power between the SeqDP and LP solvers is larger than in other months, owing to the discretization of power levels. In Figure~\ref{fig:Relative_reduction_new}, it is seen that the SeqDP reduced peak power by up to 93\% and costs by up to 94\% relative to uncontrolled charging. On average, costs fell by about 80\% and peak power decreased by approximately 78\%. Although there is a slight variation in peak grid power, the total costs obtained by LP solver and SeqDP are close, as shown in Figure~\ref{fig:Smart_uncontrolled_cost}.

\begin{figure}
	\centering
	\begin{subfigure}[b]{0.45\textwidth}
		\centering
		\includegraphics[width=\linewidth]{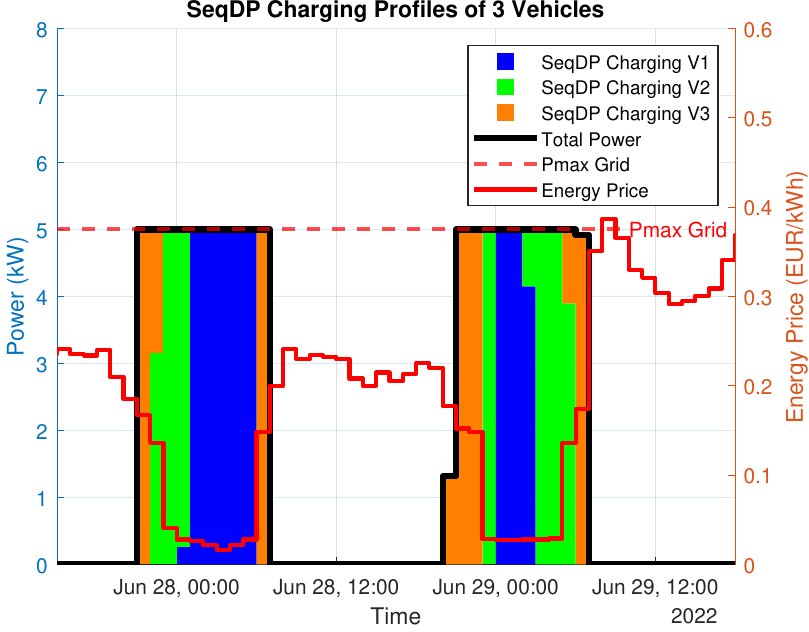}
		\caption{SeqDP smart charging profile.}
		\label{fig:charging_compare_sdp}
	\end{subfigure}
	\hfill
	\begin{subfigure}[b]{0.45\textwidth}
		\centering
		\includegraphics[width=\linewidth]{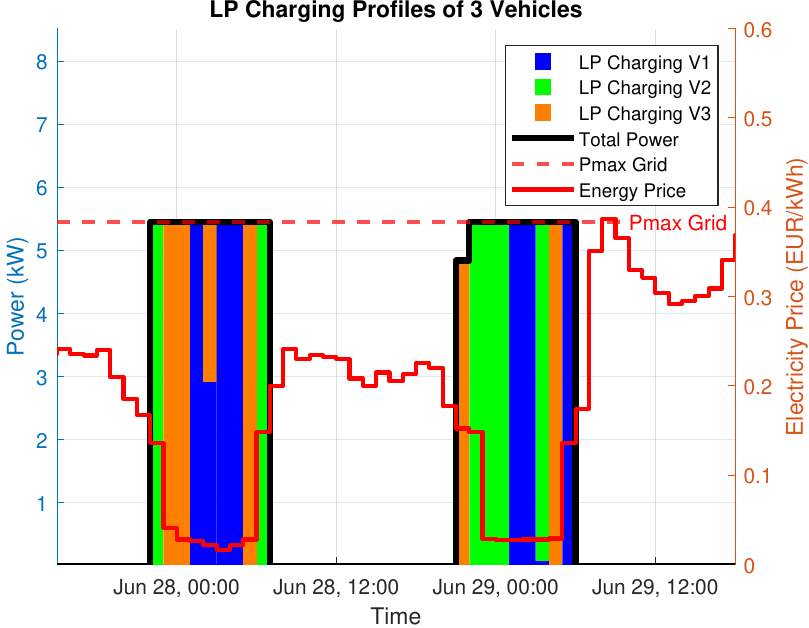}
		\caption{LP solver smart charging profile.}
		\label{fig:charging_compare_milp}
	\end{subfigure}

	\caption{SeqDP and LP solver smart charging profiles.}
	\label{fig:charging_compare}
\end{figure}

\begin{figure*}
	\centering
	\includegraphics[width=0.75\textwidth]{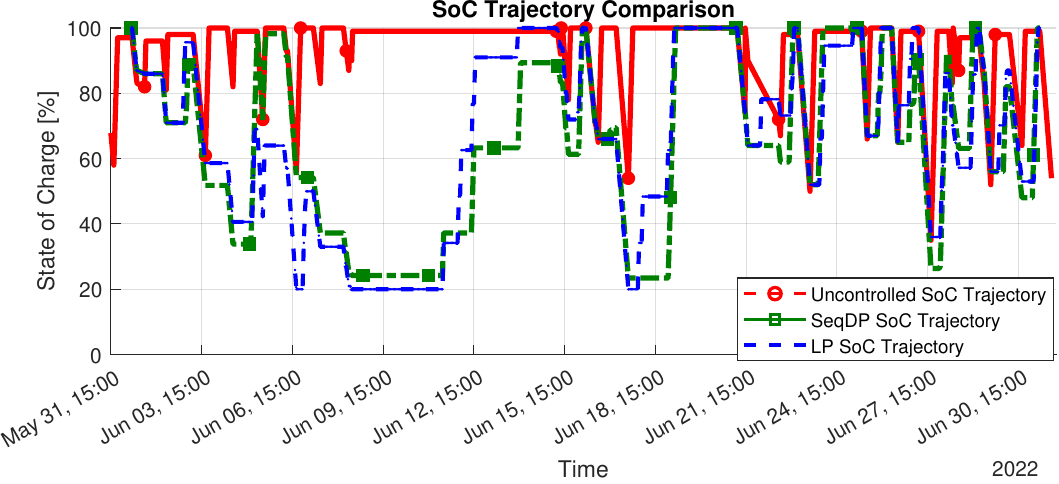}
	\caption{SoC trajectory comparison of the different charging strategies.}
	\label{fig:SOC_cmp}
\end{figure*}

In all months, both optimization methods effectively shaved power peaks and drastically reduced charging costs. The primary differences in peak power and cost arise from the charging sequences produced by each algorithm, see Figure~\ref{fig:charging_compare}. In this figure, charging is performed at the lowest energy prices, without exceeding the peak power threshold.  Additionally, Figure~\ref{fig:SOC_cmp} presents a comparison of SoC trajectories for both optimization methods versus the uncontrolled charging profile. The LP solver plan exhibits more frequent SoC states near the lower and upper bounds than the DP structure, due to vehicle charging coordination and the absence of discretization, unlike SeqDP. When the minimum state of charge is increased from 20\% to 30\%, the total monthly cost for June increases slightly from 40.25 to 40.58 EUR in LP formulation, while maintaining the same peak power of 5.45\,kW. The SeqDP result rose from 41.14 to 41.24 EUR, while the maximum grid power increased from 5 to 6 kW. With the same SoC constraint applied to the March charging plan, the LP structure increases cost by 3.05 EUR, while DP rises by 2.96 EUR.  Although this adjustment results in only a slightly higher cost increase in the LP problem solver, the outcome remains marginally better than the SeqDP method. This gap persists because state aggregation and coordination were not fully addressed in the SeqDP-based optimization in this study. However, the SeqDP’s computational speed is approximately twice that of LP solver.

\subsection{Large-scale Fleet Experiment}

\begin{figure}
	\centering
	\includegraphics[width=0.47\textwidth]{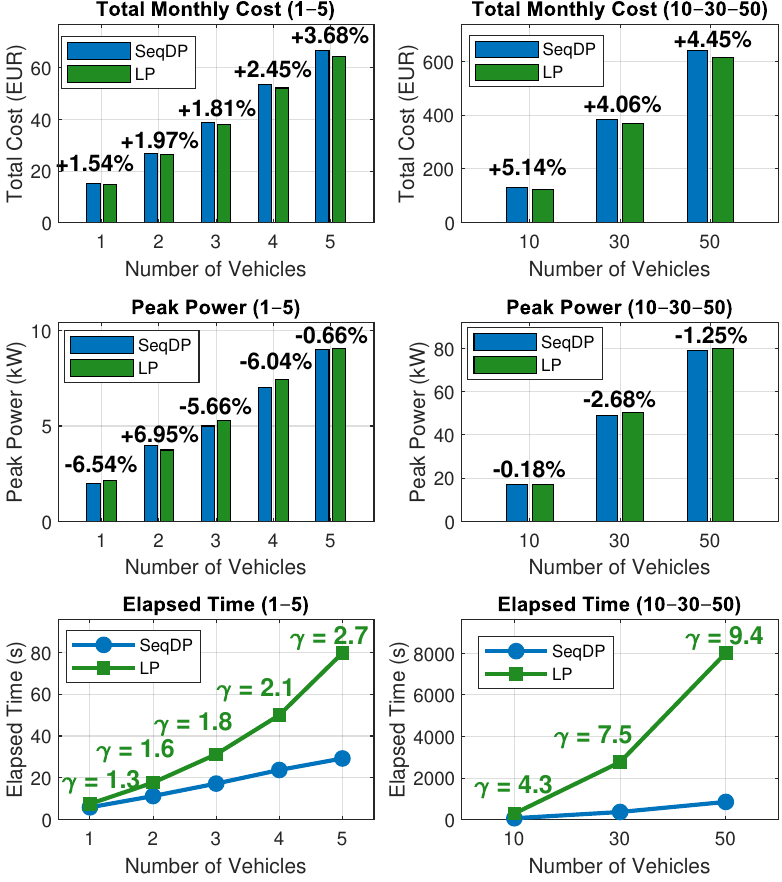}
	\caption{Evaluation of both SeqDP and LP solver methods with respect to the increasing number of vehicles in the fleet in June 2022.}
	\label{fig:veh_scaling}
\end{figure}

The large-scale fleet design comprises charging planning for up to 50 vehicles. The charging planning encompasses shuttles and accounts for different operating horizons and energy requirements. Figure~\ref{fig:veh_scaling} presents the performance comparison of the two optimization methods, the LP solver and the proposed SeqDP approach, under increasing fleet sizes during June 2022. Both smart planning methods' results are compared in terms of total energy cost and time complexity. Let $\gamma$ denote the ratio of LP solver to SeqDP computational time. For 50 shuttles, LP solver requires 9.4 times the SeqDP's solve time; for 30 and 10 vehicles, the slowdowns are 7.5 and 4.3 times, respectively. Thus, the SeqDP consistently exhibits improved computational efficiency across the shown scales, and the LP solver’s nonlinearity complexity indicates that runtimes become significantly longer if the fleet size increases.
Relative to the LP solver benchmark, the SeqDP solution shows deviations in cost and peak power. These percentage differences are annotated above the cost and power bars. In all cases, the relative cost deviation remains below $6\%$. However, this is only one case, and the algorithm is case sensitive.

\begin{figure}
	\centering
	\begin{subfigure}[b]{0.45\textwidth}
		\centering
		\includegraphics[width=\linewidth]{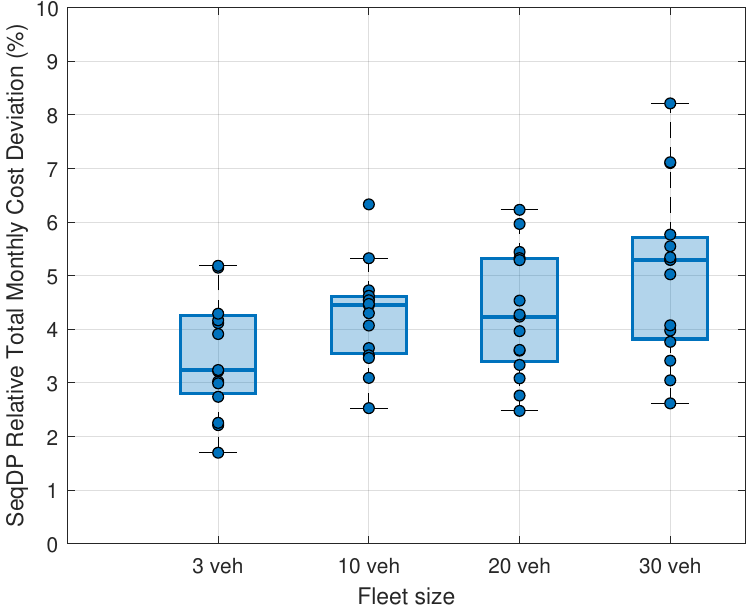}
		\caption{Relative total cost deviation of SeqDP with respect to LP solver.}
		\label{fig:veh_diff_rand_10}
	\end{subfigure}
	\hfill
	\begin{subfigure}[b]{0.45\textwidth}
		\centering
		\includegraphics[width=\linewidth]{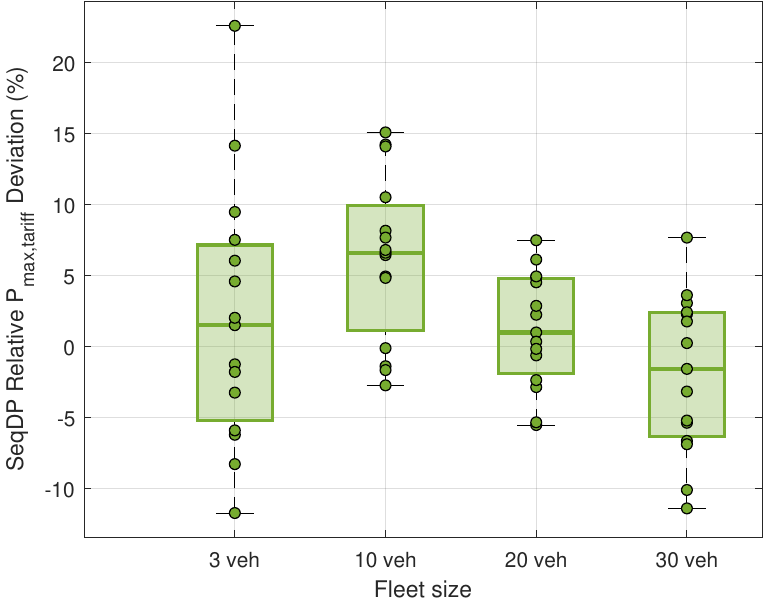}
		\caption{Relative maximum power deviation of SeqDP with respect to LP solver.}
		\label{fig:veh_diff_rand_20}
	\end{subfigure}

	\caption{Different randomization of operational schedules for fleets of 3, 10, 20, and 30 vehicles. (Both methods show more than a 90\% reduction in cost and power compared to uncontrolled charging.)}
	\label{fig:veh_diff_rand_all}
\end{figure}

Figure~\ref{fig:veh_diff_rand_all} presents 15 randomized scenarios of the operational schedules for a 3, 10, 20, and 30-vehicle fleet to test the robustness of the algorithm. In every case, the SeqDP’s relative cost remains under 9\%, on average approximately 4.5\% relative to the LP formulation. In contrast, the relative peak grid power can increase by up to 22\% due to the lack of vehicle coordination. Regarding their median metrics, the relative total cost and the relative maximum grid power are 4.24\% and 2.88\%, respectively, for the 20 vehicles case.  In the 10-vehicle case,  the medians of relative total cost and relative maximum grid power are 4.45\% and 6.62\%, while in the 30-vehicle case, they are 5.29\% and 3.63\%, respectively. For the small fleets, a 1 kW deviation from the optimal 10 kW power can result in a relative error of up to 10\%. Because the peak power is discretized to integers, this effect is more likely to be observed here. In the 30-vehicle scenario, when available charging slots are shorter or more energy is required for the limited time slot, uncoordinated charging tends to push peak power higher, resulting in relatively high errors in peak power needed to sustain each vehicle's operation. Nevertheless, according to Figure~\ref{fig:veh_diff_rand_all}, the major distribution of relative error in the total cost is lower than 6\%.

The results clearly demonstrate that while the LP solver achieves higher accuracy and produces globally optimal solutions, its computational time increases exponentially with the number of vehicles due to the growth in state space. In contrast, the proposed SeqDP method exhibits slight nonlinearity on the time axis, making it significantly more scalable. Despite its simplified structure, SeqDP delivers results that closely approximate the global optimum, offering a favorable trade-off between accuracy and computational efficiency.

While the computational time for the classical DP scales linearly with the horizon $T$, the observed runtime of the SeqDP is not strictly linear. Complexity increases with the number of decision slots and is further affected by shared-grid coupling among vehicles. With $K$ vehicles and $T$ time steps, feasibility at each $(k,t)$ requires evaluation of the aggregate prior load $\sum_{\nu<k} P_\nu(t)$. Enumerating vehicles in sequence implies that, at a fixed time $t$, vehicle $1$ requires $0$ prior terms, vehicle $2$ requires $1$, \dots, and vehicle $K$ requires $K-1$, yielding quadratic, $\mathcal{O}(K^2)$, increase across the horizon. Constraining the outer search over $P_{\text{max,tariff}}$ to fixed windows (e.g., $[0,15]$ kW or $[35,50]$ kW) reduces constant factors but does not eliminate the coupling term, so the runtime remains nonlinear in $K$. In contrast, Gurobi’s LP treats all time periods jointly.

\section{Conclusion}
\label{sec:conclusion}

This paper focuses on an efficient optimization strategy for the smart charging of electrified shuttle fleets over a monthly time horizon, driven by recent power tariffs in order to avoid higher peak loads, localized imbalances, transformer overload risk, and increased charging costs. The classical dynamic programming method exhibits exponential computational complexity as the number of states increases, making it especially impractical for month-long analyses aimed at minimizing fleet-wide peak demand. To control computational growth, a heuristic method, SeqDP, is adopted to enable systematic aggregation and efficient sequential computation, rather than joint optimization of an LP problem. Contrary to the prevailing literature, the proposed SeqDP algorithm evaluates each shuttle individually. It assesses feasibility against its own timetable, even over a one-month horizon, without battery aggregation or vehicle clustering. Due to operational constraints, the model incorporates discrete energy states, time-varying tariffs, and state-of-charge targets to achieve a scalable, reduced-peak-power, and cost-effective approach.  The methodology is also applied to case studies for both small and large fleets. Regarding accuracy metrics, the heuristic sequential DP approach is benchmarked against Gurobi, a widely used solver in academia and industry.

The simulation results demonstrate that smart charging strategies, particularly those based on optimization techniques such as Sequential Dynamic Programming (SeqDP) and LP solver, offer substantial advantages over uncontrolled charging profiles in terms of cost efficiency and grid peak reduction. Both methods effectively minimize charging costs and shave monthly peak power demands, with LP formulation solver achieving slightly better performance due to its global optimization capabilities and multi-vehicle coordination. However, SeqDP remains a practical and scalable alternative, achieving near-optimal results with significantly lower computational complexity. Considering the results, the SeqDP achieves cost and power reductions of more than 90\% compared to the uncontrolled charging strategies.

Overall, while the LP solver serves as a benchmark for optimal performance, SeqDP emerges as a computationally efficient approach that delivers comparable results in large-scale applications. For 50 shuttles, the LP’s solve time is $9.4$ times that of the SeqDP; for 30 and 10 vehicles, the ratios are $7.5$ and $4.3$, respectively. Given the growth in nonlinear complexity with problem size, the LP solver’s runtime is expected to become impractical for larger fleets.  The results show that the SeqDP is consistently faster across the tested cases.

\printcredits

\bibliographystyle{cas-model2-names}

\bibliography{cas-refs}

\end{document}